\documentclass[pre,byrevtex,amsmath,amssymb,preprint]{revtex4}

\usepackage{graphicx}
\usepackage{dcolumn}
\usepackage{bm}

\begin{document}
\preprint{PREPRINT}

\title[Short Title]{Polyelectrolyte Multilayering on a Charged Planar Surface}

\author{Ren\'e Messina}
\email{messina@thphy.uni-duesseldorf.de}
\affiliation
{Institut f\"ur Theoretische Physik II,
Heinrich-Heine-Universit\"at D\"usseldorf,
Universit\"atsstrasse 1,
D-40225 D\"usseldorf,
Germany}

\date{\today}

\begin{abstract}
  The adsorption of highly \textit{oppositely} charged flexible
  polyelectrolytes (PEs) on a charged planar substrate is investigated by means of 
  Monte Carlo (MC) simulations. 
  We study in detail the equilibrium structure of the first few PE layers.
  The influence of the chain length and of a (extra) non-electrostatic short 
  range attraction between the polycations and the negatively charged substrate is considered.
  We show that the stability as well as the microstructure of the PE layers 
  are especially sensitive to the strength of this latter  interaction.
  Qualitative agreement is reached with some recent experiments.
\end{abstract}

\maketitle

\section{Introduction}

PE multilayer structures are often obtained in a 
so-called layer-by-layer method by alternating exposure of a 
charged substrate to solutions of polycations (PCs) and polyanions (PAs).  
This widely used technique was first introduced by Decher and coworkers
\cite{Decher_1992,Decher_1997}, and its simplicity and versatility
trigger a large interest in the engineering community.
As examples of technological applications, one can mention:
biosensing \cite{Caruso_Langmuir_1998}, catalysis \cite{Onda_1999}, 
non-linear optical devices \cite{Wu_JACS_1999}, nanoparticles coating
\cite{Caruso_Science_1998,Caruso_JPCB_2001}, etc. 
%

On the theoretical side, there exit a few analytical works about PE multilayers
on charged planar surfaces based on different levels of approximation 
\cite{Solis_JCP_1999,Netz_Macromol_1999b,Castelnovo_2000}.
Solis and Olvera de la Cruz considered the conditions under which the spontaneous 
formation of polyelectrolyte layered structures can be induced by a charged 
wall \cite{Solis_JCP_1999}. 
Based on Debye-H\"uckel approximations for the electrostatic interactions
and including some lateral correlations by the consideration of given 
adsorbed PE structures,
Netz and Joanny\cite{Netz_Macromol_1999b} found a remarkable
stability of the (semi-flexible) PE multilayers supported by scaling laws.
For weakly charged flexible polyelectrolytes at high ionic
strength qualitative agreements between theory
\cite{Castelnovo_2000}, also based on scaling laws, and experimental
observations \cite{Ladam_Langmuir_2000} 
(such as the predicted thickness and net charge of the PE multilayer)  have been provided.

The important driving force for all these PE multilayering processes is of
electrostatic origin. More precisely, it is based on an overcharging mechanism, 
where the first layer overcharges the substrate and, along the PE multilayering process, 
the top layer overcharges the adsorbed PE layers underneath.  
Nevertheless, the strong correlations existing between oppositely charged
polyions, especially for highly charged PEs, provide a formidable challenge
for the understanding of the PE multilayer microstructures.
In this respect, numerical simulations are of great help. 
It is only recently, that MC simulations were carried out to study such PE structures
built up on \textit{spherical} charged surfaces \cite{Messina_Langmuir_2003}.
%

In this paper, we provide a detailed study of the PE multilayer structure
adsorbed on a charged planar surface and discuss the basic mechanisms that are 
involved there by means of MC simulations.
Our paper is organized as follows: Sec.  \ref{ Sec.simu_method}
is devoted to the description of our MC simulation technique.  
The measured quantities are specified in Sec. \ref{Sec.Target}.  
The PE monolayering is studied in Sec. \ref{Sec.monolayer}, and the PE
bilayering in Sec. \ref{Sec.bilayer}.  
Then the PE multilayering process is addressed in Sec. \ref{Sec.Multilayer}.  
Finally, Sec. \ref{Sec.conclu} contains some brief concluding remarks.

\section{Simulation details
 \label{ Sec.simu_method}}

The setup of the system under consideration is similar to that recently
investigated with a spherical substrate \cite{Messina_Langmuir_2003}. 
Within the framework of the primitive model we consider a PE
solution near a charged hard wall with an implicit solvent (water)
of relative dielectric permittivity $\epsilon_{r}\approx 80$.  
This charged substrate located at $z=0$ is characterized by a 
negative surface bare charge density $-\sigma_0 e$, where $e$ is the
(positive) elementary charge and $\sigma_0>0$ is the number of charges per unit area. 
Electroneutrality is always ensured by the presence of explicit monovalent 
($Z_c=1$) plate's counterions of diameter $a$.
PE chains ($N_+$ PCs and $N_-$ PAs) 
are made up of $N_m$ \textit{monovalent} monomers ($Z_{m}=1$) of diameter $a$. 
Hence, all microions are monovalent: $Z=Z_c=Z_m=1$.
For the sake of simplicity, we only consider here symmetrical complexes where 
PC and PA chains have the same length and carry the same charge in absolute value.  

%
%

All these particles making up the system are confined in a $L \times L \times  \tau$ box.
Periodic conditions are applied in the $(x,y)$ directions, whereas hard walls
are present at $z=0$ (location of the charged plate) and $z=\tau$ 
(location of an \textit{uncharged} wall).
To avoid the appearance of image charges \cite{Torrie_JCP_1982,Messina_image_2002}, 
we assume that on both parts of the charged plate (at $z=0$) 
the dielectric constants are the same.

The total energy of interaction of the system can be written as

\begin{eqnarray}
\label{eq.U_tot}
U_{tot} & = &  
\sum_i  \left[ U_{hs}^{(plate)}(z_i) + U_{coul}^{(plate)}(z_i) + U_{vdw}^{(plate)}(z_i) \right] + 
\\ \nonumber
&& \sum _{i,i<j} \left[U_{hs}(r_{ij}) + U_{coul}(r_{ij}) + U_{FENE}(r_{ij}) + U_{LJ}(r_{ij}) \right],
\end{eqnarray}
where the first (single) sum stems from the interaction between an ion $i$ 
(located at $z=z_i$) and the charged plate, 
and the second (double) sum stems from the pair interaction between ions $i$ and $j$
with $r_{ij}=|{\bf r}_i - {\bf r}_j|$.
All these contributions to $U_{tot}$ in Eq. (\ref{eq.U_tot})
are described in detail below.

Excluded volume interactions are modeled via a hardcore potential
defined as follows
%
\begin{equation}
\label{eq.U_hs}
U_{hs}(r_{ij})=\left\{
\begin{array}{ll}
\infty,
& \mathrm{for}~r_{ij} < a \\
0,
& \mathrm{for}~r_{ij} \geq a
\end{array}
\right.
\end{equation}
%
for the microion-microion one,
except for the monomer-monomer one \cite{note_HS}, 
and
%
\begin{equation}
\label{eq.U_hs_plate}
U_{hs}^{(plate)}(z_i)=\left\{
\begin{array}{ll}
\infty,
& \mathrm{for} \quad z_i < a/2\\
\infty,
& \mathrm{for} \quad z_i > \tau - a/2\\
0,
& \mathrm{for} \quad a/2 \leq z_i \leq  \tau - a/2
\end{array}
\right.
\end{equation}
%
for the plate-microion one.

The electrostatic energy of interaction between two ions $i$ and $j$ reads
%
\begin{equation}
\label{eq.U_coul} 
\frac{U_{coul}(r_{ij})}{k_BT} =
\pm  \frac{l_B}{r_{ij}},
\end{equation}
%
where +(-) applies to charges of the same (opposite) sign, and
$l_{B}=e^{2}/4\pi \epsilon _{0}\epsilon _{r}k_{B}T$ is the Bjerrum
length corresponding to the distance at which two monovalent ions
interact with $k_B T$.  
The electrostatic energy of interaction between an ion $i$ and the
(uniformly) charged plate reads
%
\begin{equation}
\label{eq.U_coul_plate} 
\frac{U_{coul}^{(plate)}(z_i)}{k_BT} =
\pm  2 \pi l_B \sigma_0  z_i 
\end{equation}
%
where +(-) applies to positively (negatively) charged ions.
An appropriate and efficient modified Lekner sum was utilized to compute 
the electrostatic interactions with periodicity in \textit{two} 
directions \cite{Brodka_MolPhys_2002,note_Lekner}. 
To link our simulation parameters
to experimental units and room temperature ($T=298$K) we choose
$a =4.25$ \AA\ leading to the Bjerrum length of water
$l_{B}=1.68a =7.14$ \AA.
The surface charge density of the planar macroion was chosen as 
$-\sigma_0 e \approx -0.165~ {\rm C/m^2}$.

The polyelectrolyte chain connectivity is modeled by employing a standard
FENE potential in good solvent (see, e.g., \cite{Kremer_FENE_1993}),
which reads
%
\begin{equation}
\label{eq.U_fene}
U_{FENE}(r)=
\left\{ \begin{array}{ll}
\displaystyle -\frac{1}{2}\kappa R^{2}_{0}\ln \left[ 1-\frac{r^{2}}{R_{0}^{2}}\right] ,
& \textrm{for} \quad r < R_0 \\ \\
\displaystyle \infty ,
& \textrm{for} \quad r \geq R_0 \\
\end{array}
\right.
\end{equation}
%
with $\kappa = 27k_{B}T/ a^2$ and $R_{0}=1.5 a$.
The excluded volume interaction between chain monomers is taken into
account via a purely repulsive Lennard-Jones (LJ) potential given
by
%
\begin{equation}
\label{eq.LJ}
U_{LJ}(r)=
\left\{ \begin{array}{ll}
\displaystyle
4\epsilon \left[ \left(\frac{a}{r}\right)^{12}
-\left( \frac{a}{r}\right) ^{6}\right] +\epsilon,
& \textrm{for} \quad r \leq 2^{1/6} a \\ \\
0,
& \textrm{for} \quad  r > 2^{1/6} a
\end{array}
\right.
\end{equation}
%
where $\epsilon=k_BT$.
These parameter values lead to an equilibrium bond length $ l=0.98a$.

An important interaction in PE multilayering is the \textit{non}-electrostatic 
\textit{short ranged attraction}, $U_{vdw}^{(plate)}$, between the planar macroion and the PC chain.  
To include this kind of interaction, we choose without loss of
generality a (microscopic) van der Waals (VDW) potential of interaction between
the planar macroion and a PC monomer that is given by
%
\begin{equation}
\label{eq.U_vdw}
U_{vdw}^{(plate)}(z)=-\epsilon \chi_{vdw}
\left( \frac{a}{z+a/2} \right)^6
\hspace{0.5cm} \textrm{for}~ z \geq a/2 ,
\end{equation}
%
where $\chi_{vdw}$ is a positive dimensionless parameter describing the
strength of this attraction. Thereby, at contact (i.e., $z=a/2$),
the magnitude of the attraction is $\chi_{vdw} \epsilon=\chi_{vdw} k_BT$ which
is, in fact, the relevant characteristic of this potential.
Since it is not straightforward to directly link this strength of
adsorption  to experimental values, we choose $\chi_{vdw}=5$ 
(also considered among other values in the case of a spherical macroion \cite{Messina_Langmuir_2003}), 
so as to mimic good ``anchoring'' properties to the planar substrate.

All the simulation parameters are gathered in Table \ref{tab.simu-param}.
The set of simulated systems can be found in
Table \ref{tab.simu-runs}.
The equilibrium properties of our model system were obtained by using standard canonical MC 
simulations following the Metropolis scheme \cite{Metropolis_JCP_1953,Allen_book_1987}.
Single-particle moves were considered with an acceptance ratio of
$30\%$ for the monomers and $50\%$ for the counterions.
Typically, about $5 \times 10^4$ to $\times 10^6$ MC steps per
particle were required for equilibration, and about $5 \times
10^5 - 10^6$ subsequent MC steps were used to perform measurements. 
To improve the computational efficiency, we omitted the presence of PE counterions
when $N_+=N_-$ so that the system is still globally electroneutral. 
We have systematically checked for $N_+=N_-=20$ (system $C$) that
the (average) PE configurations (especially the monomer distribution) are indistinguishable, 
within the statistical uncertainty, from those where PE counterions are explicitly 
taken into account, as it should be.

\section{Measured quantities
 \label{Sec.Target}}

We briefly describe the different observables that are going to be measured.  
In order to characterize the PE adsorption, we compute the monomer density
$n_{\pm}(z)$ that is normalized as follows

\begin{equation}
\label{eq.n_r}
\int ^{\tau-a/2}_{a/2} n_\pm(z) L^2 dz = N_\pm N_m
\end{equation}
%
where $+(-)$ applies to PCs (PAs). This quantity is of special interest
to characterize the degree of ordering in the vicinity of the planar macroion surface.

The total number of  accumulated monomers $\bar{N}_{\pm}(r)$ within a distance $z$
from the planar macroion is given by
%
\begin{equation}
\label{eq.N_r}
\bar{N}_\pm(z) = \int ^{z}_{a/2} n_\pm(z') L^2 dz'
\end{equation}
%
where $+(-)$ applies to PCs (PAs). 
This observable will be addressed in the study of  PE monolayer
(Sec. \ref{Sec.monolayer}) and  PE bilayer (Sec. \ref{Sec.bilayer}).

Another relevant quantity is the global \textit{net fluid
charge} $\sigma(z)$ which is defined as follows
\begin{equation}
\label{Eq.Qr}
\sigma(z)=\int ^{z}_{a/2} \left[
\tilde{n}_{+}(z')-\tilde{n}_{-}(z')\right] dz',
\end{equation}
%
where $\tilde{n}_+$ ($\tilde{n}_-$) stand for the density of all the
positive (negative) microions (i.e., monomers and counterions).  
Thus, $\sigma(z)$ corresponds to the
net fluid charge per unit area (omitting the bare macroion surface-charge $-\sigma_0$)
within a distance $z$ from the charged wall. At the uncharged
wall, electroneutrality imposes $\sigma(z=\tau-a/2)=\sigma_0$.  
By simple application of the Gauss' law,
$\left[ \sigma(z)-\sigma_0\right]$ is directly proportional
to the mean electric field at $z$.  Therefore $\sigma(z)$ can
measure the \textit{screening} strength of the macroion-plate charge
by the neighboring solute charged species.

\section{Monolayer
 \label{Sec.monolayer}}

In this part, we study the adsorption of PC chains
for two chain lengths $N_m=10$ (system $A$) and $N_m=20$ (system $B$),
and for two different couplings $\chi_{vdw}=0$ and $\chi_{vdw}=5$.
Experimentally, this would correspond to the formation of the
\textit{first} polyelectrolyte layer.
This is a decisive step to elucidate the even more complex PE multilayer structures
where additionally PAs are also present.
 
Here, where $N_-=0$ (i.e., no polyanions), global electroneutrality is ensured 
by the presence of explicit PC's counterions (i.e., monovalent anions) and 
the macroion-plate's counterions (i.e., monovalent cations). 
Also, we recall that the total number of monomers, $N_+N_m=200$, is identical for both 
systems $A$ and $B$ under consideration (see Table \ref{tab.simu-runs}). 
Hence, the total monomer charge is the same for systems $A$ and $B$.

\begin{center}
\begin{figure}
\includegraphics[width = 12.0 cm]{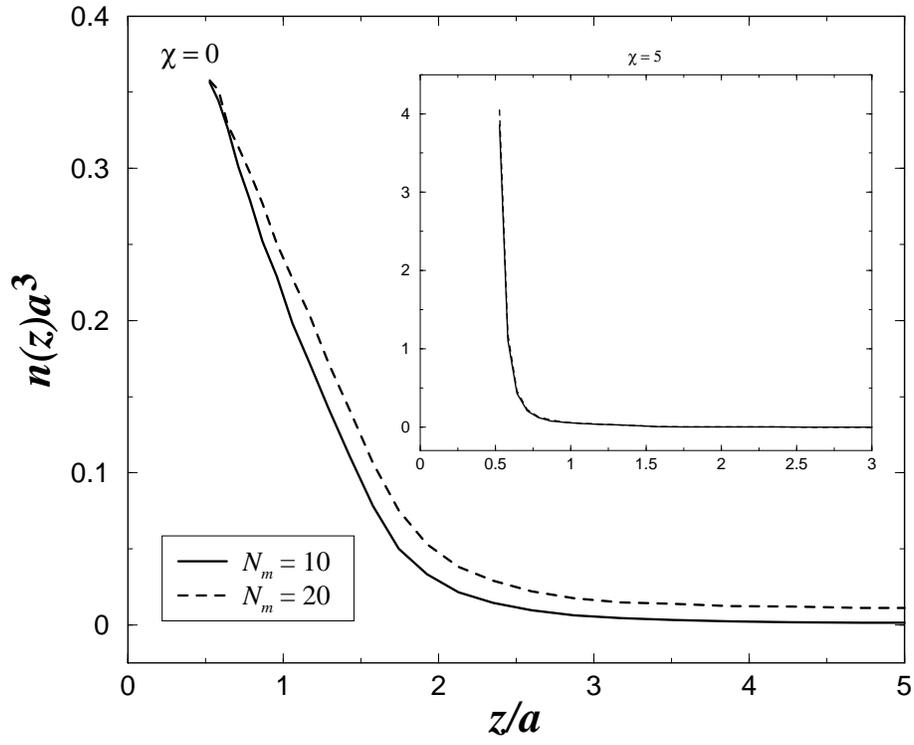}
\caption{Messina, Macromolecules}
\label{fig.nz_monolayer}
\end{figure}
\end{center}

The profiles of the monomer density $n_+(z)$ are depicted in Fig. \ref{fig.nz_monolayer}. 
At $\chi_{vdw}=0$, the density $n_+(z)$ near contact ($z \sim a/2$)
is basically independent on the chain size $N_m$. But away from the surface, the density of monomers
is slightly higher for larger $N_m$. This is a combined effect of (i) entropy and 
(ii) electrostatic correlations.
These underlying mechanisms at $\chi_{vdw}=0$ can be explained with simple ideas as follows:
\begin{itemize}
\item At fixed number of total monomers, entropic effects are larger the shorter the chains, 
and in the limiting case of $N_m=1$ (i.e., the electrolyte limit) entropy effects are maximal leading
to the highest monomer ``release''. 
It is to say that the chain connectivity lowers the entropy of the system.
\item  In parallel, electrostatic correlations 
\cite{Shklowskii_PRE_1999b,Messina_PRE_2001,Levin_RepProgPhys_2002} are also higher 
the higher the valence of the adsorbed particles. 
In our case $N_m$ plays the role of the polyion valence. 
\end{itemize}
The density $n_+(z)$ near contact increases considerably with $\chi_{vdw}$ 
(here about one order of magnitude) as expected. 
It turns out that with $\chi_{vdw}=5$, the $n_+(z)$-profiles are basically 
identical for $N_m=10$ and $N_m=20$.
This is due to the sufficiently strong non-electrostatic attractive force that can
overcompensate the antagonistic entropic effects that were more efficient at $\chi_{vdw}=0$.

The fraction $\bar{N}_+(z)/(N_+ N_m)$  of adsorbed monomers can be found in Fig. \ref{fig.Nz_monolayer}. 
At a $z$-distance of $1.5a$ from the planar macroion
surface (corresponding to a width of two monomers), about 90\% of the monomers are
adsorbed for $\chi_{vdw}=5$ against only $\sim 50\%$ for $\chi_{vdw}=0$. 
Again, at given $\chi_{vdw}$, $\bar{N}_+(z)/(N_+ N_m)$ is larger for longer chains due to the 
same coupled effects of entropy and electrostatic correlations explained above.

The (global) net fluid charge $\sigma(z)$ is reported in Fig. \ref{fig.Qz_monolayer}. 
In all cases we observe a macroion-surface charge reversal (i.e., $\sigma(z)/\sigma_0>1$).
The position $z=z^*$ at which $\sigma(z^*)$ gets
its maximal value decreases with $\chi_{vdw}$, due to the $\chi_{vdw}$-enhanced
adsorption of the PCs.  Concomitantly, this \textit{overcharging} increases with
$\chi_{vdw}$, since the (extra) gain in energy by macroion-monomer VDW interactions
can better overcome (the higher $\chi_{vdw}$) the cost of the self-energy
stemming from the adsorbed excess charge \cite{Messina_Langmuir_2003}.  
More quantitatively, we have $\sigma(z^*)/\sigma_0 \approx 1.7$ at  $\chi_{vdw}=5$ against only
$\sigma(z^*)/\sigma_0 \approx 1.25$ at $\chi_{vdw}=0$.
Note that the maximal value of charge reversal of $(200-90)/90=122\%$ 
(i.e., $\sigma(z^*)/\sigma_0=2.22$)
allowed by the total charge of PCs can not be reached due to a slight accumulation of microanions.
In agreement with the profiles of $n_+(z)$ and $\bar{N}_+(z)$ (see Fig. \ref{fig.nz_monolayer} and 
Fig. \ref{fig.Nz_monolayer}), at given  $\chi_{vdw}$, the overcharging gets higher the higher 
the chain length. 
Those (locally) overcharged states should be the driving force for the building of subsequent PE bilayers when
PA chains are added.

\begin{center}
\begin{figure}
\includegraphics[width = 16.0 cm]{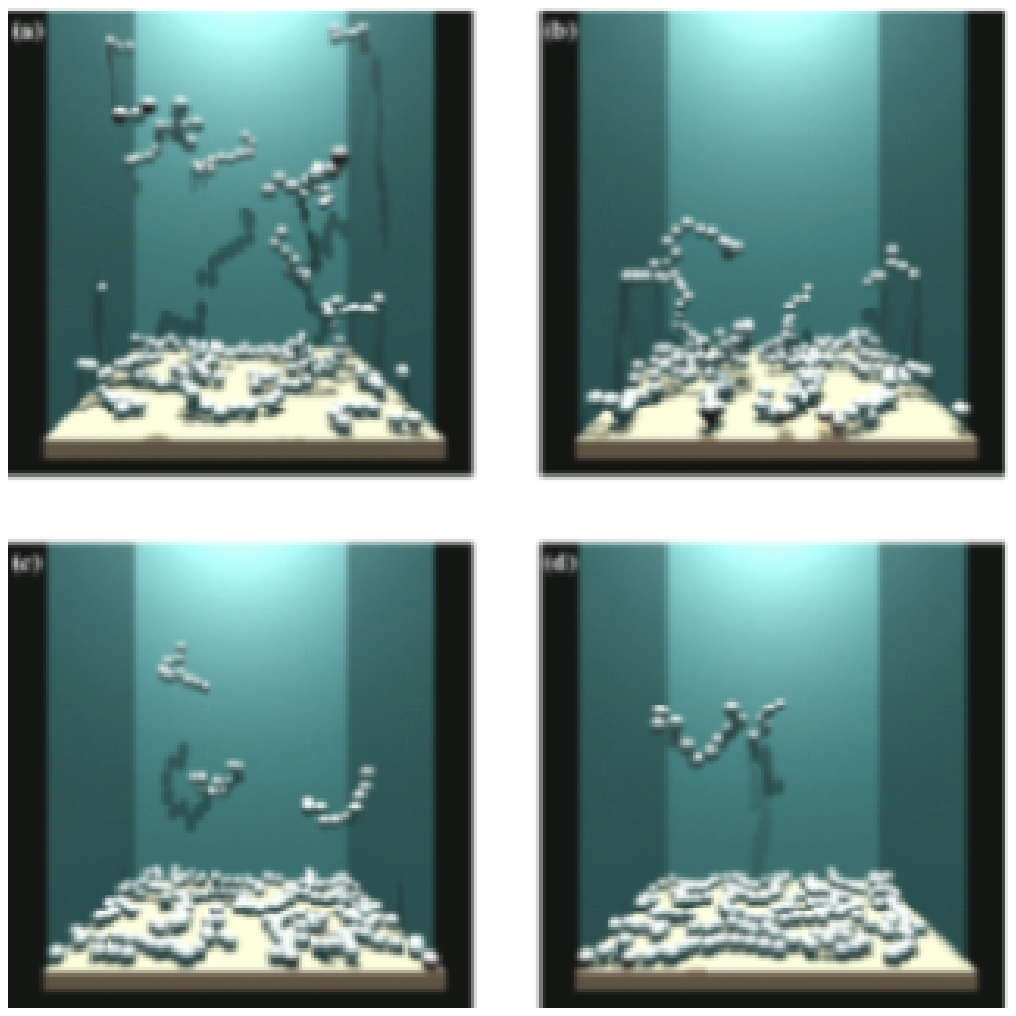}
\caption{Messina, Macromolecules}
\label{fig.snap_monolayer}
\end{figure}
\end{center}

Typical equilibrium configurations can be found in Fig. \ref{fig.snap_monolayer}. 
The qualitative difference between $\chi_{vdw}=0$ [Figs. \ref{fig.snap_monolayer}(a) and (b)]
and $\chi_{vdw}=5$  [Figs. \ref{fig.snap_monolayer}(c) and (d)] is rather spectacular.
Without additional VDW attraction ($\chi_{vdw}=0$) the adsorption is much weaker
than at $\chi_{vdw}=5$, where in the latter situation the $z$-fluctuation is very weak within the
adsorbed layer. Basically the first layer is glued at $\chi_{vdw}=5$, and the excess PC chains float 
in the solution. It is typically this type of configurations for the first layer that is wanted 
in experimental situations. 

The next section (Sec. \ref{Sec.bilayer}) that concerns bilayering will show 
that the (enhanced) stability of this first layer is decisive for the onset 
of multilayers.
%
\section{Bilayer
 \label{Sec.bilayer}}

We now consider the case where additionally PA chains are present (systems $C$ and $D$),
so that we have a neutral polyelectrolyte complex (i.e., $N_+N_m=N_-N_m=200$).
Global electroneutrality is ensured by the counterions of the planar macroion as usual.
For such parameters, the final equilibrium structure consists essentially of bilayers
with sometimes the onset of a weakly stable third layer.
Experimentally this would correspond to the process of the \textit{second}
polyelectrolyte layer formation (with system $A$ or $B$ as the initial state).  
We stress the fact that this process is fully reversible for the parameters
investigated in our present study. In particular, we checked that the same
final \textit{equilibrium} configuration is obtained either by (i) starting
from system $A$ or $B$ and then adding PAs or (ii) starting directly 
with the mixture of oppositely charged polyelectrolytes.

\begin{center}
\begin{figure}
\includegraphics[width = 10.0 cm]{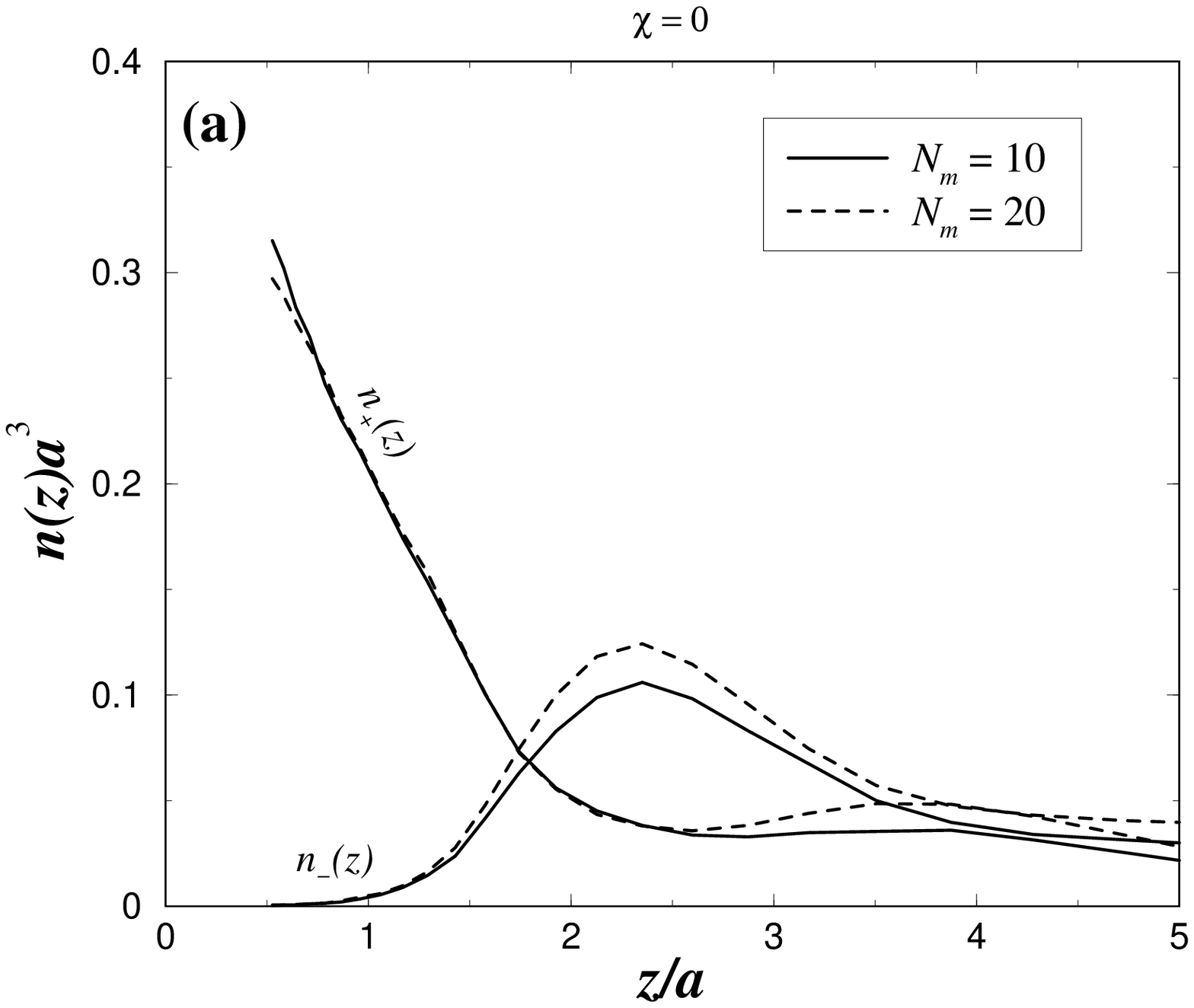}
\includegraphics[width = 10.0 cm]{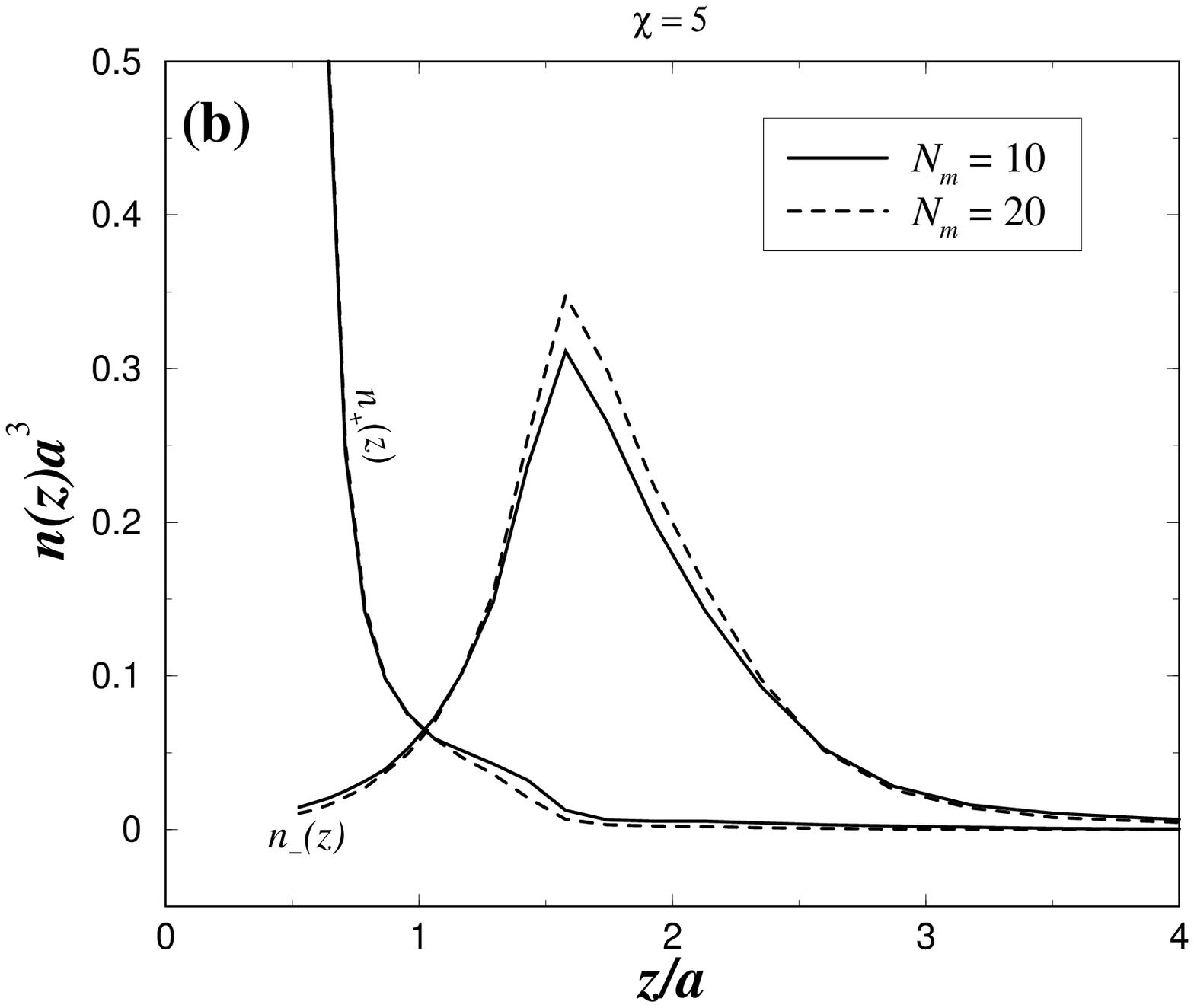}
\caption{Messina, Macromolecules}
\label{fig.nz_bilayer}
\end{figure}
\end{center}

The profiles of the monomer density $n_\pm(z)$ at $\chi_{vdw}=0$ and $\chi_{vdw}=5$ are
depicted in Fig. \ref{fig.nz_bilayer}(a) and Fig. \ref{fig.nz_bilayer}(b), respectively.
The corresponding microstructures are sketched in Fig. \ref{fig.snap_bilayer}.  
The density of PC monomers $n_+(r)$ near contact increases 
considerably with $\chi_{vdw}$ as expected.  
Interestingly, at $\chi_{vdw}=0$, a comparison with systems $A$ and $B$ 
(see Fig. \ref{fig.nz_monolayer})
indicates that the adsorption of PC monomers is weaker when additional PAs are present. 
This effect was already observed with spherical substrates \cite{Messina_Langmuir_2003},
and the same mechanisms apply here to planar surfaces.
More explicitly,  the PC chain tends to build up a globular state 
(reminiscent of the classical \textit{bulk} PE collapse \cite{Hayashi_JCP_2002}) by getting 
complexed to the PA chain, as well illustrated in Fig. \ref{fig.snap_bilayer}(a) and
Fig. \ref{fig.snap_bilayer}(b).
Thereby, the mean monomer coordination number (or the mean number of monomer neighbors)
gets higher which is \textit{both} (i) entropically and 
(ii) energetically (at least from the PE complex viewpoint) favorable. 
This PC desorption is only appreciable at sufficiently low $\chi_{vdw}$ where the energy loss 
stemming from the PC desorption is well balanced 
(or even overcompensated depending generally on the parameters) 
by the energy gained in building a PC-PA globular structure.
This ``auto-globalization'' is also enhanced by increasing $N_m$ as it should be
[compare Fig. \ref{fig.snap_bilayer}(a) and (b)]. 
Note also that there is a small second peak in $n_+(z)$ at $z \approx 3.8a$
[see Fig. \ref{fig.nz_bilayer}(a)], which
is rather the signature of a strong PC-PA globalization than a third PE layer.
Besides, the peak in the PA density $n_-(z)$ located at $z=z^* \approx 2.3a$
[see Fig. \ref{fig.nz_bilayer}(a)], which is relatively far
from that of a compact bilayer where $z^* = 1.5a$, indicates the diffuse character of the bilayer 
at $\chi_{vdw}=0$.

At $\chi_{vdw}=5$, the $n_+(z)$-profiles are basically identical 
for $N_m=10$ and  $N_m=20$. In contrast to  $\chi_{vdw} = 0$, $n_+(z)$ near contact is somewhat larger
at $\chi_{vdw} = 5$  and it is going to be explained later in the discussion of  $\bar{N}_+(z)$.
As far as the PA density $n_-(z)$ is concerned, we see that the  peak is roughly 2-3 times higher 
(depending on $N_m$) with  $\chi_{vdw}=5$ than with  $\chi_{vdw}=0$. Also, its position 
($z^* \approx 1.5a$) corresponds to that of a compact bilayer. 
A visual inspection of Fig. \ref{fig.snap_bilayer}(c) and (d) confirms this feature. 
This again shows how important is the role of extra non-electrostatic attractive 
force for the stability of bilayers.

An intermediate conclusion can be drawn from the above findings and especially from the microstructures 
depicted in Fig. \ref{fig.snap_bilayer}:
\begin{itemize}
\item  True  bilayering (i.e., flat and dense layers) can only
       occur at \textit{non-zero} $\chi_{vdw}$, as already reported for spherical charged 
       substrates \cite{Messina_Langmuir_2003} with large curvature.
\end{itemize}
An interesting common characteristic of the microstructures at $\chi_{vdw}=0$ and $\chi_{vdw}=5$ is the formation
of \textit{small islands} (along the substrate)  made up of more or less flat (depending on $\chi_{vdw}$) PC-PA complexes,
easily identifiable at $N_m=20$ [see Fig. \ref{fig.snap_bilayer}(b) and Fig. \ref{fig.snap_bilayer}(d)].

\begin{center}
\begin{figure}
\includegraphics[width = 12.0 cm]{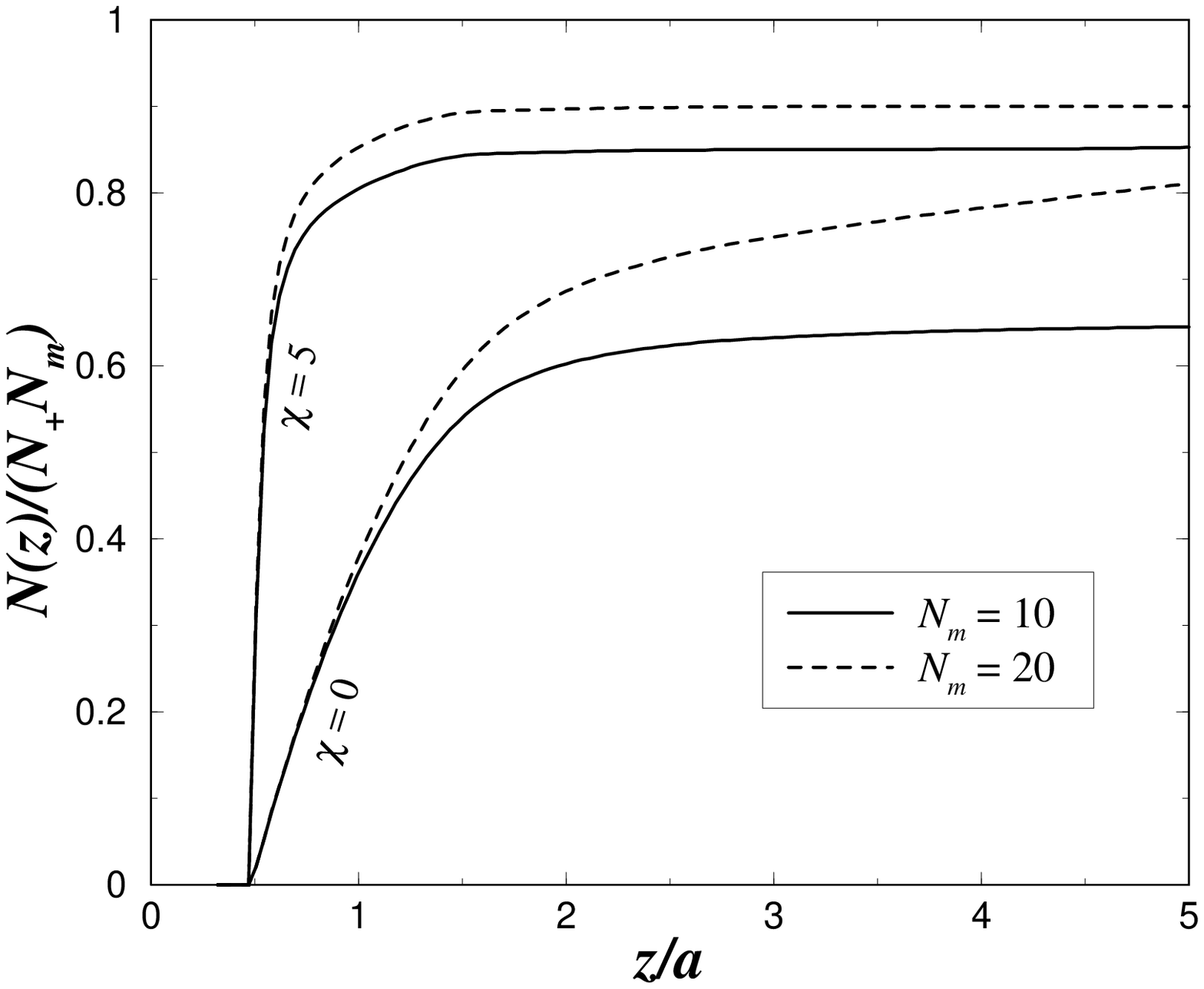}
\caption{Messina, Macromolecules}
\label{fig.Nz_monolayer}
\end{figure}
\end{center}

\begin{center}
\begin{figure}
\includegraphics[width = 10.0 cm]{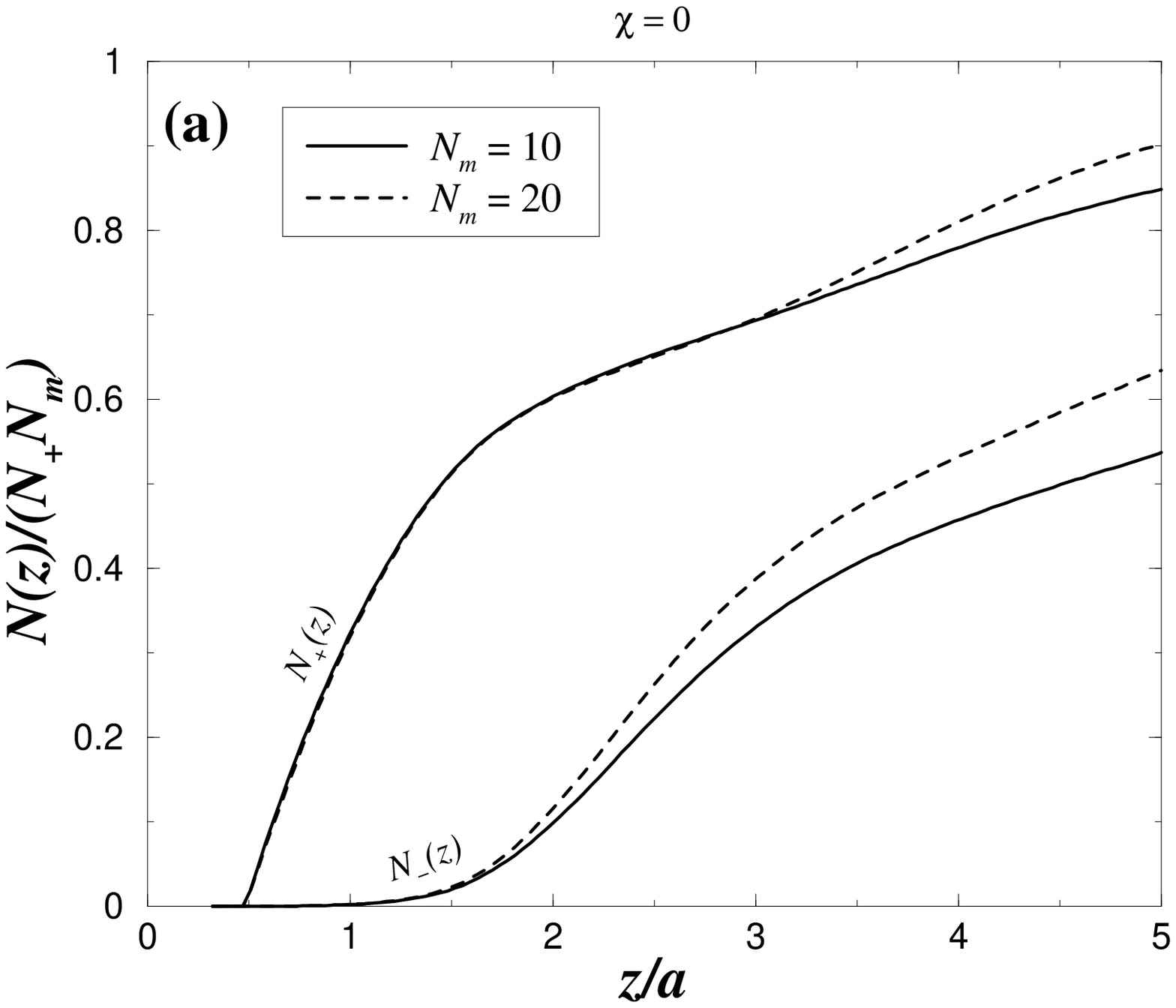}
\includegraphics[width = 10.0 cm]{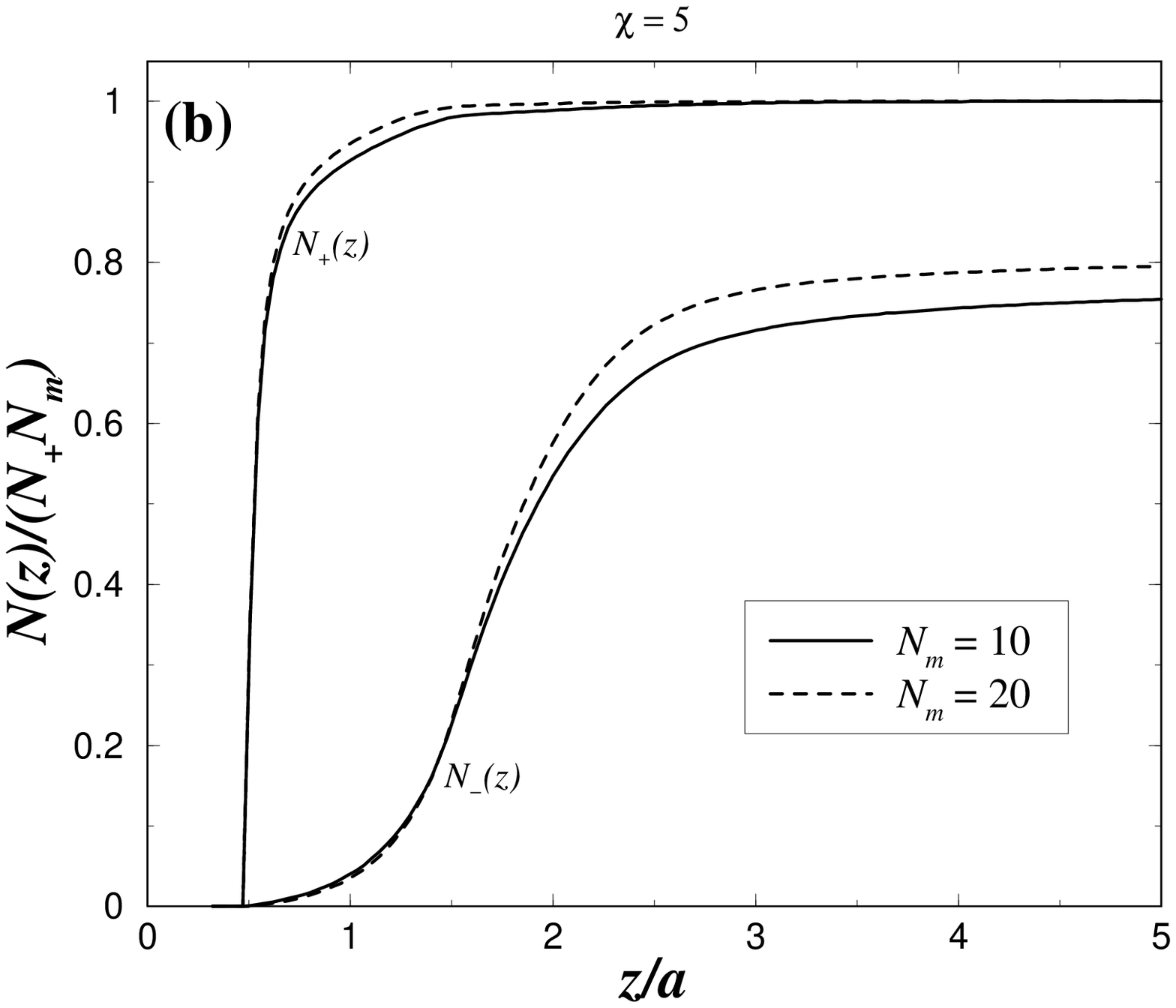}
\caption{Messina, Macromolecules}
\label{fig.Nz_bilayer}
\end{figure}
\end{center}

The fraction $\bar{N}_+(z)/(N_+ N_m)$  of adsorbed monomers at  $\chi_{vdw}=0$ and  $\chi_{vdw}=5$
can be found in Fig. \ref{fig.Nz_bilayer}(a) and (b), respectively. 
A close look at Fig. \ref{fig.Nz_bilayer}(a) reveals a smaller PC monomer accumulation (at  $\chi_{vdw}=0$) up 
to $z \approx 3a$ (independently of $N_m$) than in the case where PA chains were absent 
(compare with Fig. \ref{fig.Nz_monolayer}).
This is fully consistent with the formation of PC-PA globules (relevant at $\chi_{vdw}=0$) 
leading to the effective PC desorption already discussed above.
In parallel, this PC-PA globalization tends to cancel the effect of chain length $N_m$ on $\bar{N}_+(z)$.
On the other hand, at  $\chi_{vdw}=5$, the situation is qualitatively different where the presence of PAs
now induces an \textit{increase} of $\bar{N}_+(z)$ 
[compare Fig. \ref{fig.Nz_bilayer}(b) and Fig. \ref{fig.Nz_monolayer}].
This phenomenon can be explained by electrostatic correlation effects. 
Indeed, at $\chi_{vdw}=5$, the highly stable PC layer attracts more PA monomers than at 
$\chi_{vdw}=0$, and thereby,  "super'' dipoles  made of PC-PA monomer pairs build up, 
that are perceptible in Fig. \ref{fig.snap_bilayer}(c) and Fig. \ref{fig.snap_bilayer}(d). 
This leads to a strong attractive correlation interaction between the plate and those dipoles. 
In other terms the effect of finite $\chi_{vdw}$ is to (strongly) \textit{polarize} 
the adsorbed charged chains. 
Note also that a $\chi_{vdw}=5$ the PC-PA globalization is much less favorable 
than at $\chi_{vdw}=0$ due to the higher cost of PC desorption energy in the former case. 
As a net effect there can be more adsorbed PC monomers compared to $\chi_{vdw}=0$. 
In that case of $\chi_{vdw}=5$, it is precisely this mechanism that tends to cancel the 
effect of $N_m$ on $\bar{N}_+(z)$.
As far as the PA monomer fraction $\bar{N}_-(z)$ is concerned, Fig. \ref{fig.Nz_bilayer} 
shows that the adsorption of  monomers is much weaker and more diffuse at  $\chi_{vdw}=0$ 
than at $\chi_{vdw}=5$, as expected from Figs. \ref{fig.nz_bilayer} and \ref{fig.snap_bilayer}.

\begin{center}
\begin{figure}
\includegraphics[width = 12.0 cm]{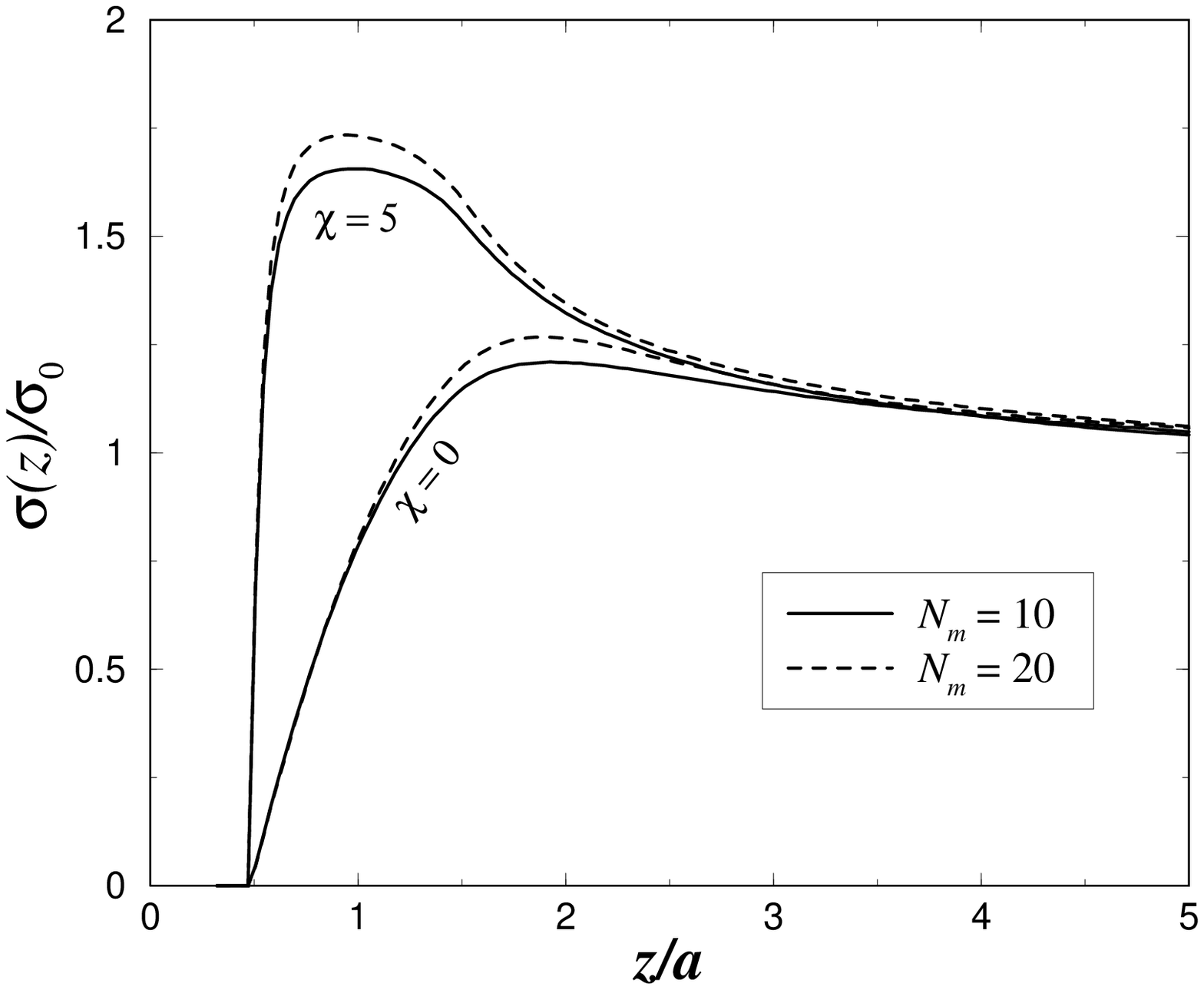}
\caption{Messina, Macromolecules}
\label{fig.Qz_monolayer}
\end{figure}
\end{center}

\begin{center}
\begin{figure}
\includegraphics[width = 12.0 cm]{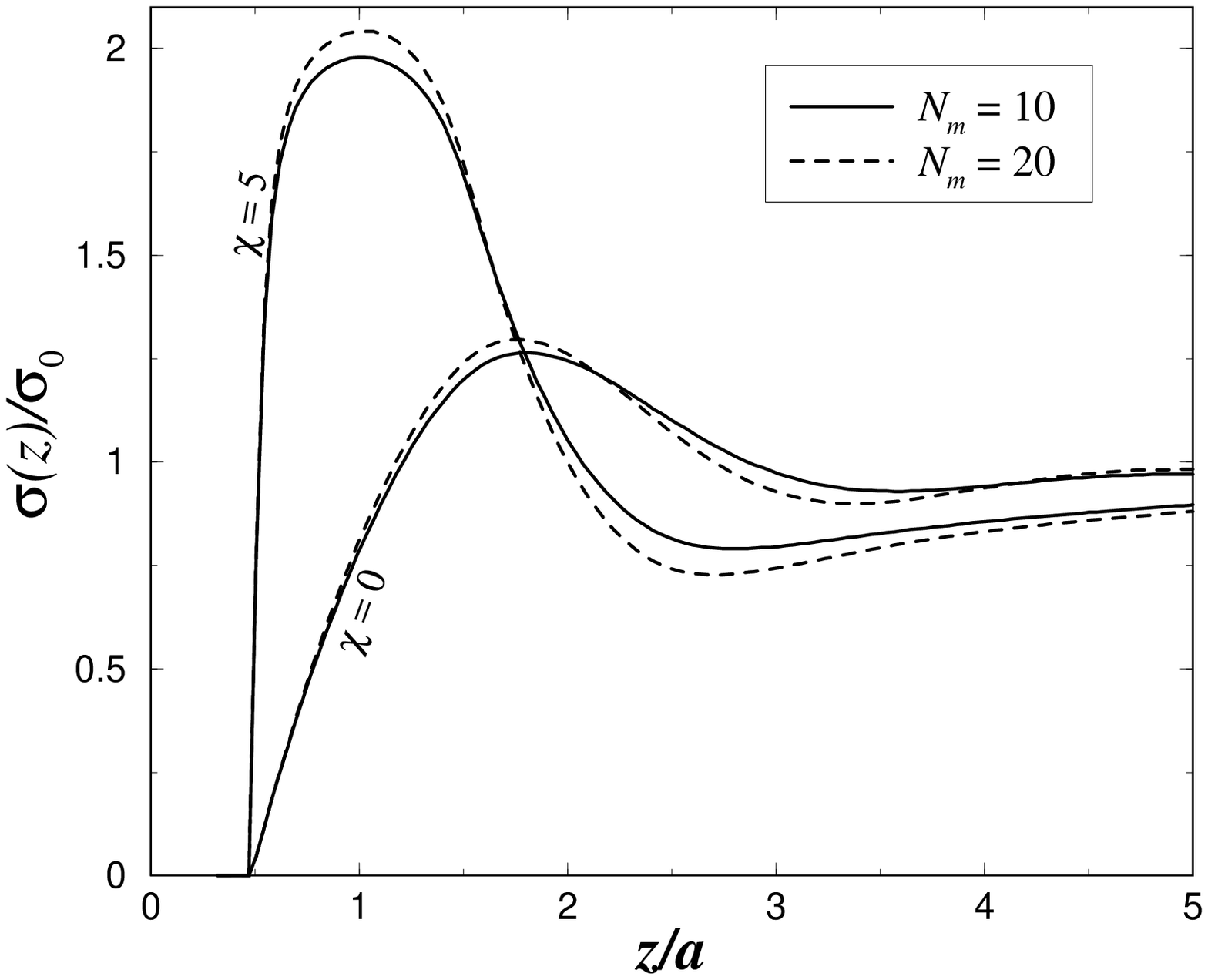}
\caption{Messina, Macromolecules}
\label{fig.Qz_bilayer}
\end{figure}
\end{center}

The net fluid charge $\sigma(z)$ is reported in Fig. \ref{fig.Qz_bilayer}.
In all cases, the planar macroion gets overcharged and undercharged as
one gets away from its surface. 
That is we have to deal with charge \textit{oscillations}.
Our results clearly show that the amplitude of those oscillations is
systematically larger at high $N_m$, as also observed without PAs (see Fig. \ref{fig.Qz_monolayer}). 
This is consistent with the idea that lateral electrostatic correlations are enhanced by 
increasing the valence of the polyions (here $N_m$).
Nevertheless, as soon as \textit{oppositely charged polyions} can interact, 
there is a subtle interplay between 
clustering and the lateral correlations of polyions that governs the degree of 
overcharging near the planar macroion.
At  $\chi_{vdw}=5$, we observe a significantly higher overcharging than without PAs 
(compare with Fig. \ref{fig.Qz_monolayer}). This is in agreement with the profiles of
$N_+(z)$ discussed previously. 
However, the positions of the first peak 
($z^* \approx a$ for $\chi_{vdw}=5$ and $z^* \approx 1.8a$ for $\chi_{vdw}=0$) 
in $\sigma(z)$ remain nearly unchanged by the presence of PAs 
(compare Fig. \ref{fig.Qz_bilayer} with Fig. \ref{fig.Qz_monolayer}).

\section{Multilayer
 \label{Sec.Multilayer}}

Presently, we consider the case where there are enough polyelectrolytes 
($N_+N_m=N_-N_m=400$) in the system to produce \textit{multilayers} (systems $E$ and $F$). 
Hence, compared to systems $C$ and $D$, we have now doubled the polyelectrolyte concentration.
Global electroneutrality is ensured by the counterions of the planar macroion as usual.

\begin{center}
\begin{figure}
\includegraphics[width = 10.0 cm]{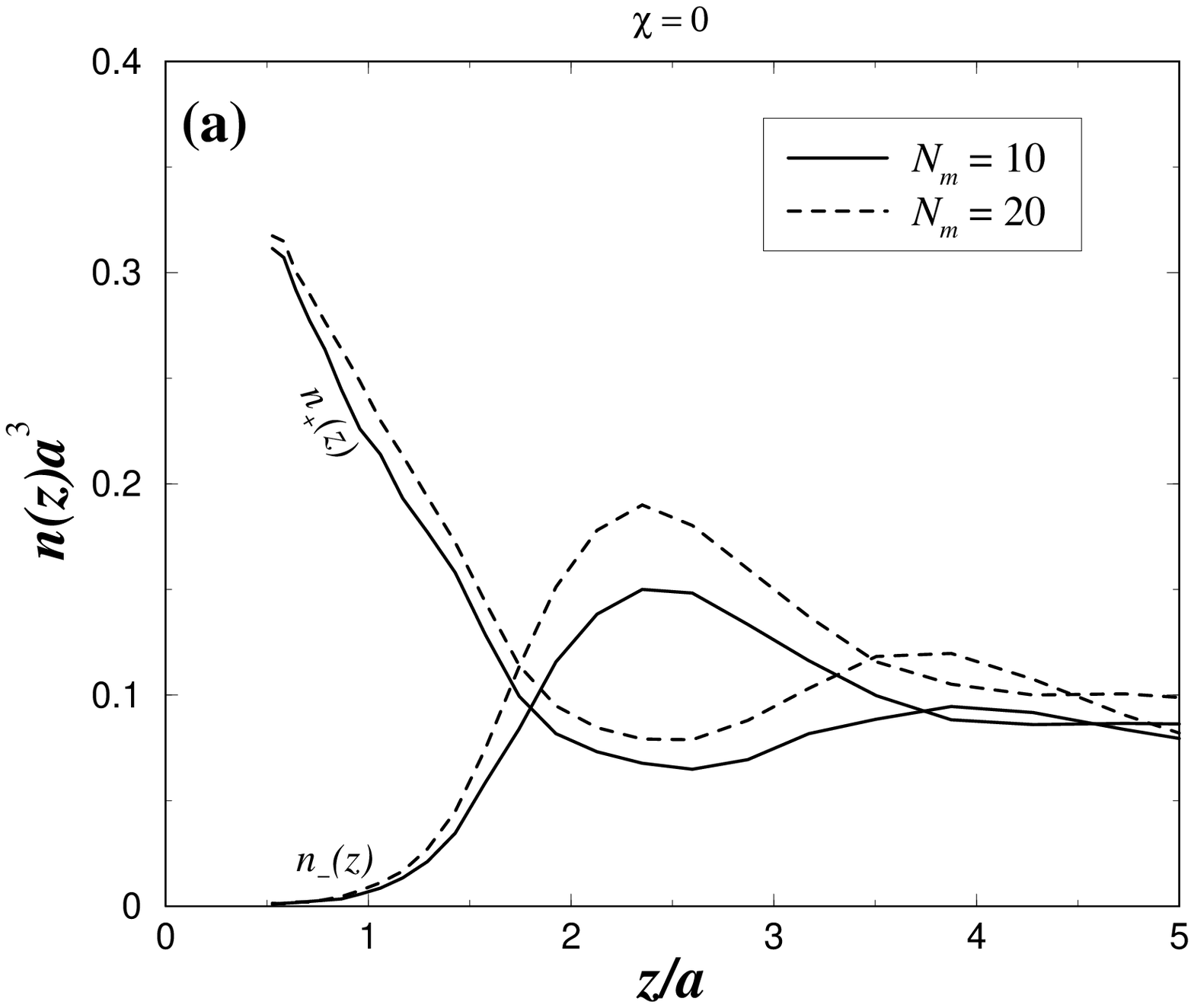}
\includegraphics[width = 10.0 cm]{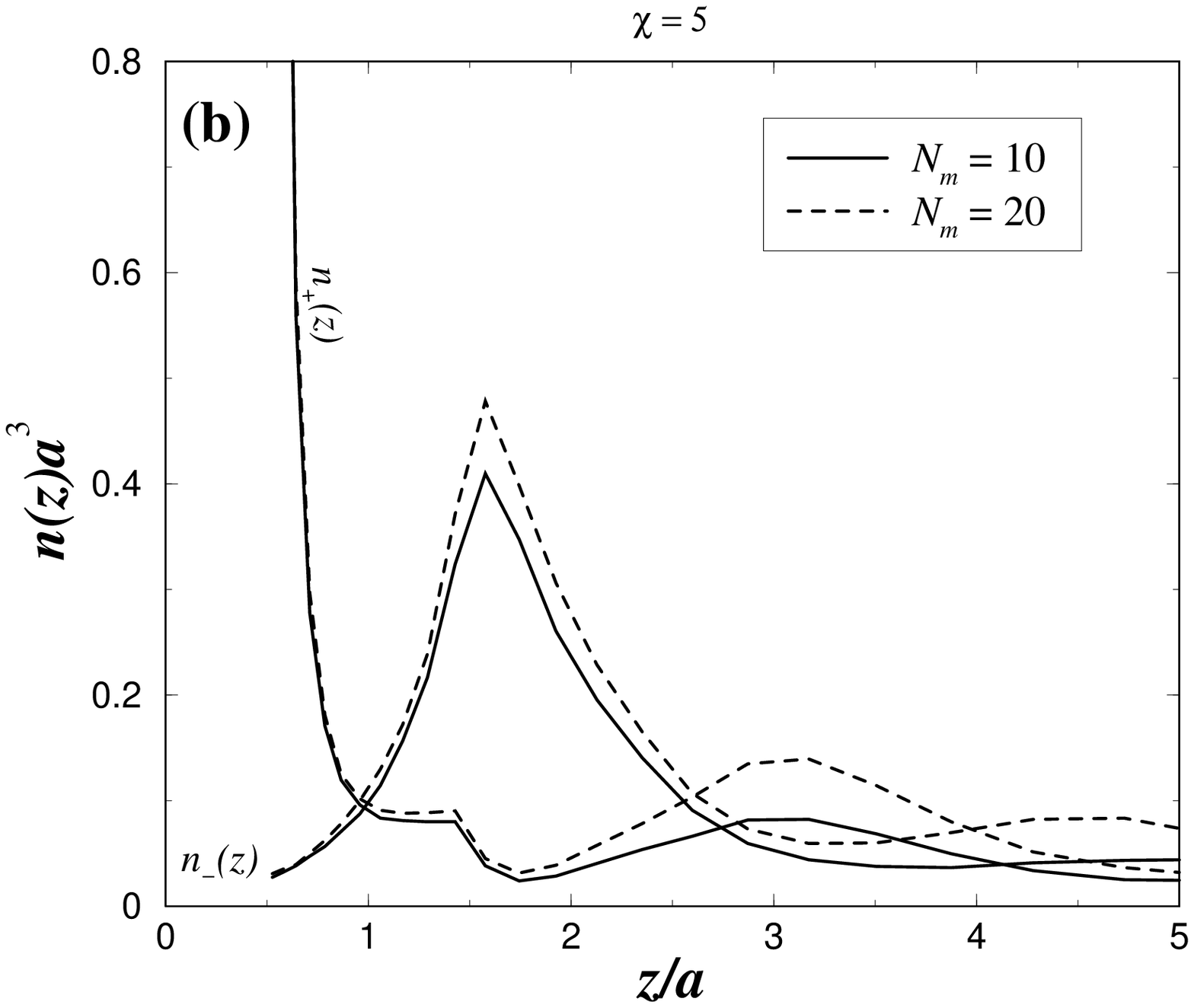}
\caption{Messina, Macromolecules}
\label{fig.nz_multilayer}
\end{figure}
\end{center}

The density profiles of $n_\pm(r)$  for $\chi_{vdw}=0$ and $\chi_{vdw}=5$ are
depicted in Fig. \ref{fig.nz_multilayer}(a) and Fig. \ref{fig.nz_multilayer}(b), respectively.
The corresponding microstructures are sketched in Fig. \ref{fig.snap_multilayer}.
In general, the densities of PC and PA monomers are systematically larger than
those found for systems $C$ and $D$ corresponding to a lower PE concentration. 
This effect is due to the fact that, at higher concentration of oppositely charged chains, 
the number of dipoles (i.e., PC-PA monomer pairs) are also larger, and from this it results 
larger plate-dipole correlations.

Even at $\chi_{vdw}=0$ with $N_m=10$, we can observe a non-negligible  second peak in
$n_+(z)$ (located at $z \approx 3.8a$) which is the signature of a third layer.  
This finding contrasts with what was observed at spherical substrates \cite{Messina_Langmuir_2003}
(also with $\chi_{vdw}=0$, $N_m=10$, and with a similar macroion surface charge density), 
where not even a stable bilayer could build up.
This radically different behavior can be accounted by geometrical arguments. 
Indeed, the potential of electrostatic interaction scales like $1/r$ in spherical geometry 
against $z$ in planar one. Hence, at sufficiently high curvature
(as it was the case in Ref. \cite{Messina_Langmuir_2003} where $N_ma/r_0 > 1$ \cite{note_curvature} 
with $r_0$ being the radius of the spherical macroion), qualitative differences from the 
planar case are then expected.
However, the corresponding microstructure [see Fig. \ref{fig.snap_multilayer}(a)] suggests
a relatively large formation of PC-PA globules leading to a  diffuse and 
porous multilayer.
Always at $\chi_{vdw}=0$, but with longer chains ($N_m=20$), Fig. \ref{fig.nz_multilayer}(a) shows
that the degree of layering is higher as expected.
This feature is well illustrated by Fig. \ref{fig.snap_multilayer}(b), where the PA monomers
are visibly more attracted to the planar macroion surface.

At $\chi_{vdw}=5$, the adsorption of monomers is drastically increased due to the
enhanced stability of the first PC layer that, in turn, induces a larger adsorption 
of the subsequent PAs and PCs.
Compared to $\chi_{vdw}=0$, all the peaks in  $n_\pm(z)$ are shifted to smaller $z$,
indicating a higher compaction.
These higher ordering and compaction at $\chi_{vdw}=5$ can be visually checked in 
Fig. \ref{fig.snap_multilayer}(c) and (d).
%
\begin{figure}
\end{figure}
%

\begin{center}
\begin{figure}
\includegraphics[width = 12.0 cm]{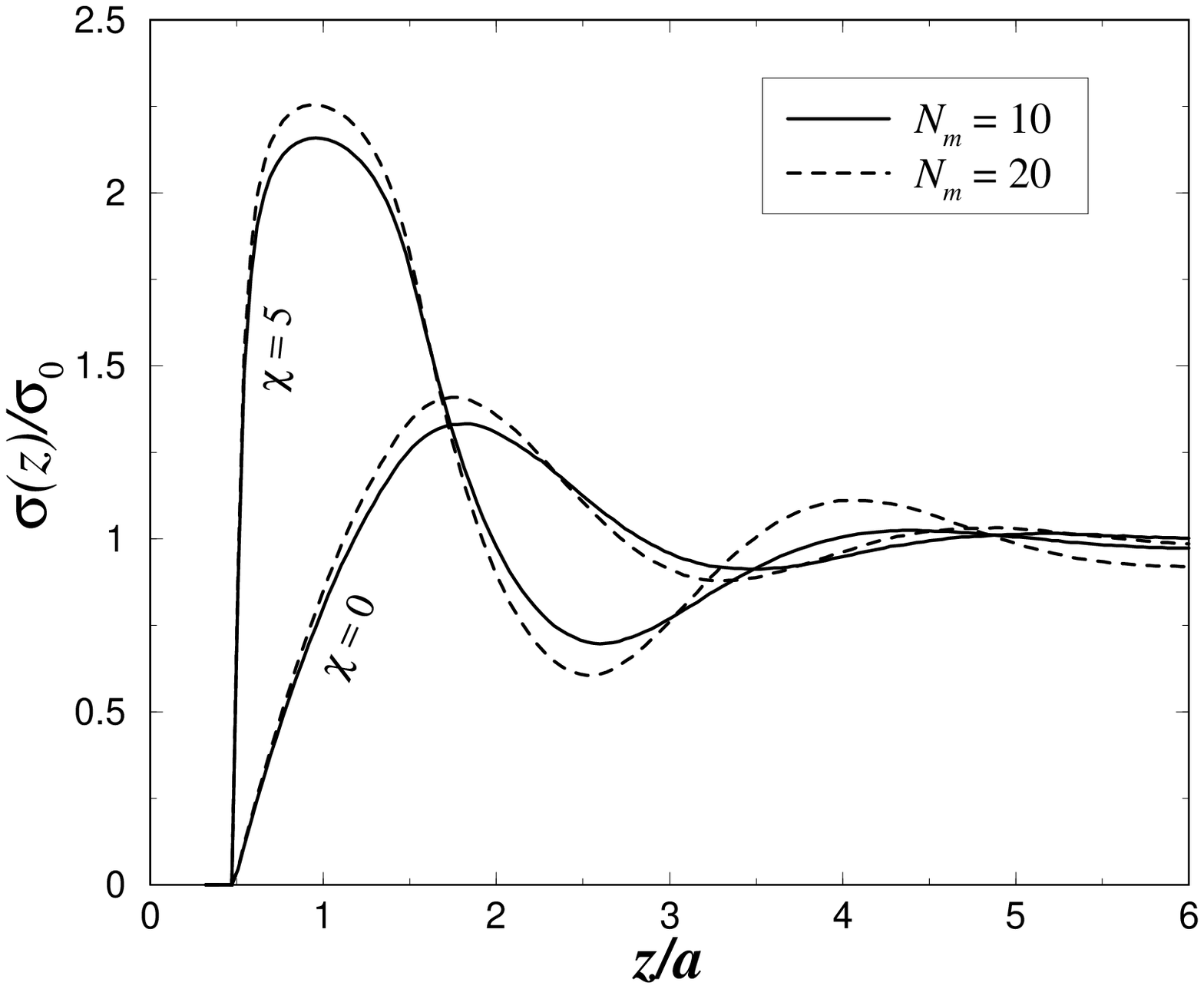}
\caption{Messina, Macromolecules}
\label{fig.Qz_multilayer}
\end{figure}
\end{center}

The net fluid charge $\sigma(z)$ is reported in Fig. \ref{fig.Qz_multilayer}.  
As expected charge oscillations are detected. 
However in all cases, the corresponding amplitude is \textit{decaying} 
(as also found in Ref. \cite{Messina_Langmuir_2003} for spherical geometry). 
Compared to the bilayer situation (see Fig. \ref{fig.Qz_bilayer}), one 
remarks that the charge oscillations are now larger due to the
enhanced ``plate-dipole'' correlations occurring at higher chain concentrations
(as discussed above). On the other hand the positions of the extrema in the charge oscillations
remain quasi unchanged.

\section{Concluding remarks
 \label{Sec.conclu}}

We first would like to briefly discuss our findings with some experimental examples.
Our results concerning the first layer (i.e., single PC layer) show that 
an additional non-electrostatic force is needed to enhance its stability.
Experimentally, this is achieved by choosing PCs with good "anchoring" properties
to a given substrate. In our model this was done by taking $\chi_{vdw}=5$.
This being said, the case $\chi_{vdw}=0$ is from a fundamental point of view interesting,
since it corresponds to a purely electrostatic regime. 

\begin{center}
\begin{figure}
\includegraphics[width = 16.0 cm]{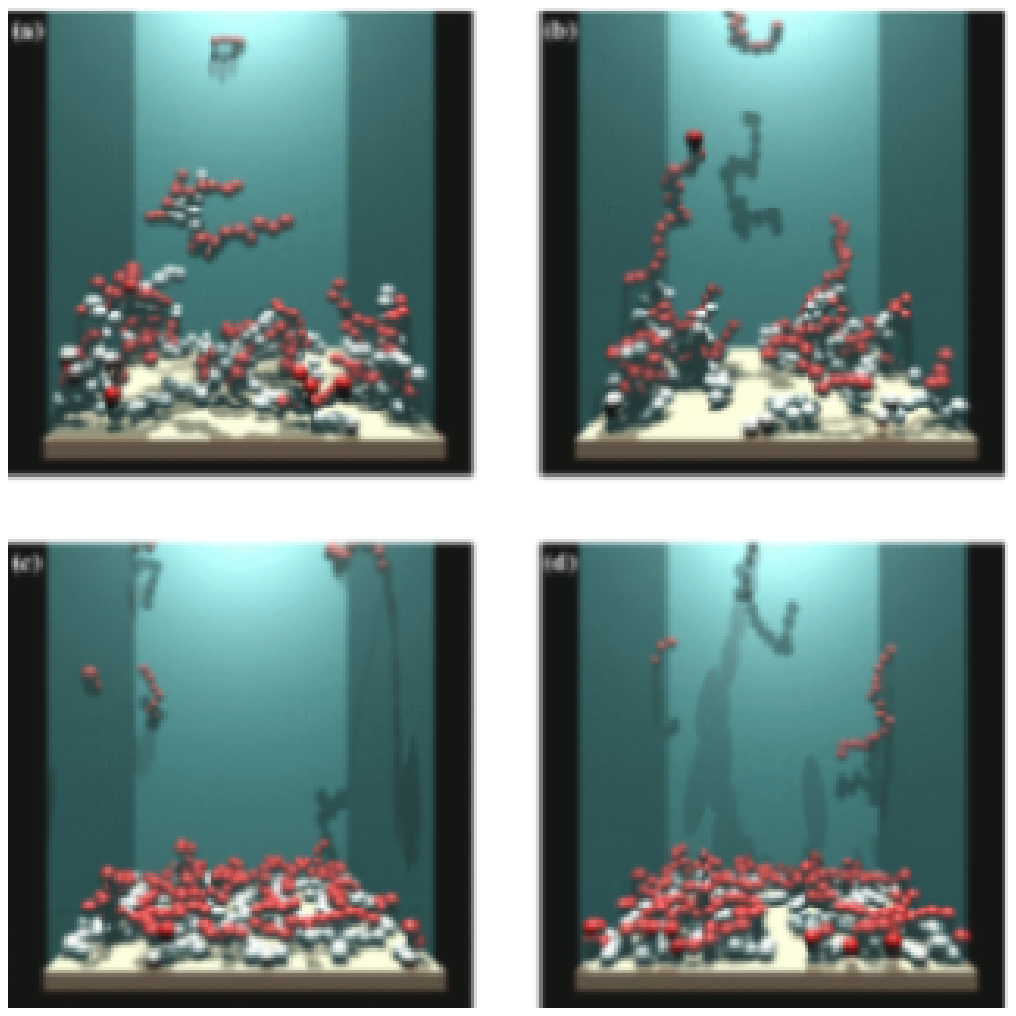}
\caption{Messina, Macromolecules}
\label{fig.snap_bilayer}
\end{figure}
\end{center}

\begin{center}
\begin{figure}
\includegraphics[width = 16.0 cm]{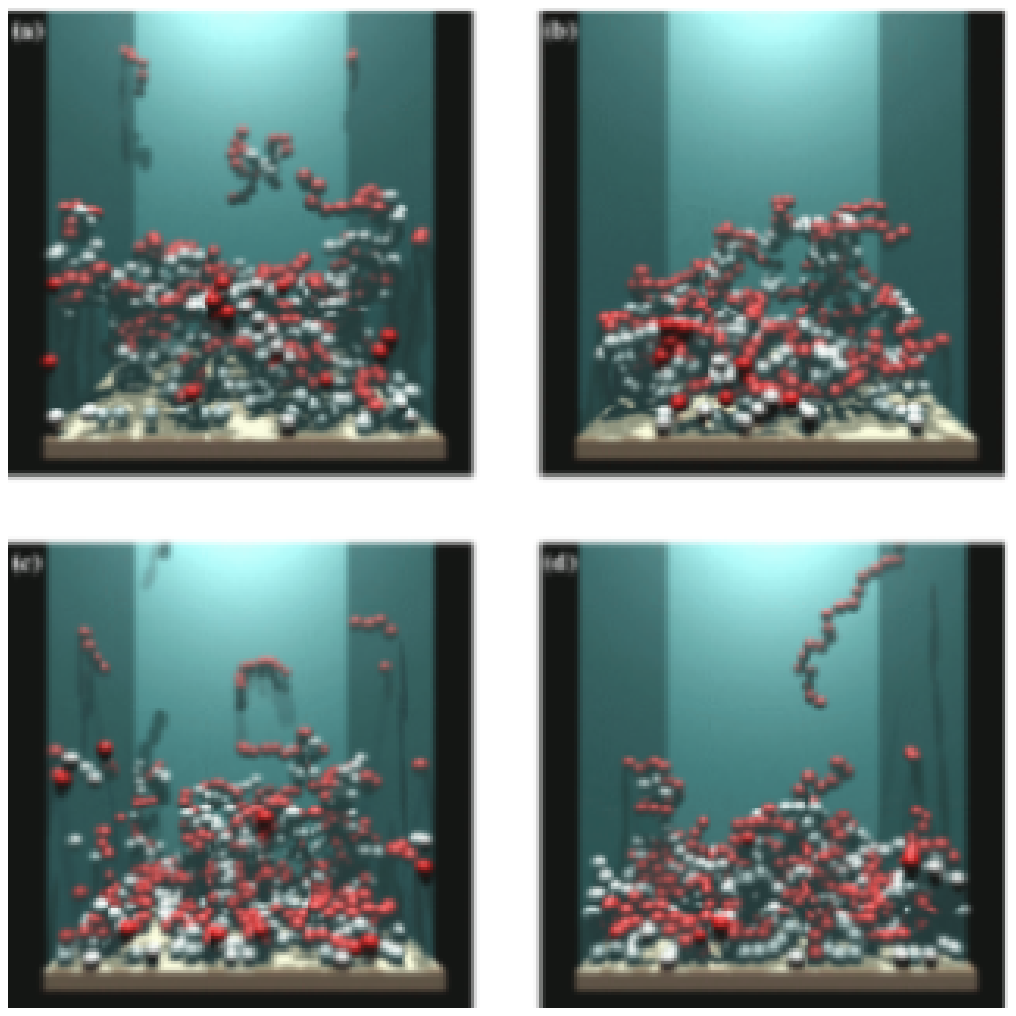}
\caption{Messina, Macromolecules}
\label{fig.snap_multilayer}
\end{figure}
\end{center}

Recently, Menchaca et al. found, by means of "liquid-cell AFM", 
that PE-complex grains appear at the first PE-layers \cite{Perez_CollSurf_2003}.
This kind of structure
(that we referred to as small islands - see Fig. \ref{fig.snap_bilayer}) are confirmed 
by our simulations. 
Concomitantly,  a significant roughness of the deposited bilayer
was also detected in this experiment, which is directly linked to the presence of those grains.
This  microstructure seems also to  be (indirectly) reported in other experiments using ellipsometry
\cite{Harris_Langmuir_2000}, 
where it is found that the structure of the two first bilayers are more porous than that of later 
bilayers. This is also in qualitative agreement with our microstructures depicted in 
Fig. \ref{fig.snap_bilayer} and Fig. \ref{fig.snap_multilayer}. 
However, more simulation data are needed to understand the 
PE structure beyond two bilayers.

The degree of charge inversion of the substrate can be indirectly obtained by measuring
the $\zeta$-potential via electrophoresis \cite{Ladam_Langmuir_2000}. 
In their experiment, Ladam et al. \cite{Ladam_Langmuir_2000} observed that, 
\textit{after a few deposited PE layers} \cite{note_thickness}, 
the $\zeta$-potential profile is symmetrically oscillating. 
This reveals a ``stationary'' regime where, successively, polycations
and polyanions are adsorbed with the same strength.
Unfortunately, it is not possible for us to investigate numerically this regime due
to the highly prohibitive computation time required there.
However, the charge oscillations observed in our $\sigma(z)$-profiles indicate that
by increasing the amount of layers, one first increases the amplitude of 
these oscillations. This confirms at least the general experimental evidence of
the non-stationary regime a the \textit{early stage} of PE multilayering.

We also would like to mention the possible effect of image charges stemming from 
the dielectric discontinuity between the substrate 
(typically $\epsilon_r \approx 2-5$) and the solvent (here $\epsilon_r \approx 80$), 
as is the case under experimental conditions.
It is expected that image forces become especially relevant for PE monolayering 
(i.e., when PCs solely are present) \cite{messina_PE_image}. 
Indeed for multivalent ions, a  strong self-image repulsion occurs and leads
to a shifted density-profile $n_+(z)$ with a maximum located somewhat further 
than the contact region \cite{Messina_image_2002,Torrie_JCP_1982}. 
However for \textit{PE multilayering}, due to the presence of oppositely charged PEs
the effect of image forces is considerably reduced 
(especially sufficiently away from the wall) due to the (self-)screening of the image 
charges. 

In summary, we have investigated by means of extensive MC simulations the
equilibrium buildup of the few first layers adsorbed on a charged planar substrate.
Two parameters were considered: (i) the chain length $N_m$ and 
(ii) the extra non-electrostatic short-range attraction
(characterized here by $\chi_{vdw}$) between the planar macroion surface and the polycation chains.

For the bilayering, it was demonstrated that, within the electrostatic regime (i.e., $\chi_{vdw}=0$),
significant PC-PA globules build up leading to a very ``porous'' and diffuse bilayer structure. 
The PC-PA globalization is enhanced with $N_m$.
At sufficiently large $\chi_{vdw}$ (here $\chi_{vdw}=5$), the bilayer is 
much less diffuse and the oppositely charged chains are more polarized, leading
to a high stability of the structure.

The same qualitatively applies to the case of the two-bilayer (i.e., four PE layers) 
adsorption.
Within this regime of layering as investigated here (up to four layers), we also
found a non-linear regime, where for instance the separation of the peaks in the monomer 
densities are not identical. 
This is in qualitative agreement with the finding of Ladam et al. \cite{Ladam_Langmuir_2000}
where they reported a non-linear regime in the so-called ``region I'' corresponding
to the PE multilayer-region close to the buffer
\cite{note_thickness}.
The effect of $N_m$ is to globally enhance the stability of the multilayer structure 
due to the higher electrostatic correlations and also due to entropic effects. 

A future study should take into account the rigidity of the chain, which
can drastically change the multilayer structure depending the stiffness.
The formation of PE multilayers on \textit{cylindrical} substrates seems also to be 
a promising research area, and to our knowledge
it has never been investigated so far \cite{messina_rod}.

\acknowledgments The author thanks F. Caruso, H. L\"owen, S. K. Mayya and E. P\'erez for 
helpful and stimulating discussions.  


\begin{thebibliography}{10}
\bibitem{note_HS}
Only the monomer-monomer excluded volume interaction was not modeled by a
hard-sphere potential. There, a purely (truncated and shifted) repulsive Lennard-Jones potential was
used [see Eq.~\eqref{eq.LJ}] so as to be compatible with the FENE potential employed
to generate the chain connectivity.
\bibitem{note_Lekner}
To further improve the computational efficiency, the corresponding
  electrostatic energies were initially stored into tables. Bear in mind, that
  such simulations (despite of this initial storage) involving Lekner sums are
  extremely time consuming. As an example, the equilibration of system $F$ with
  a satisfactory statistical accuracy requires about one month CPU time with a
  Pentium4 processor of 2.0 Ghz.
\bibitem{note_curvature}
Note that in the opposite limit of low curvature where $N_ma/r_0<< 1$, the
spherical case can be well approximated by a planar interface.
\end{thebibliography}

\providecommand{\refin}[1]{\\ \textbf{Referenced in:} #1}

\newpage
\begin{center}
{\large TABLES}
\end{center}
%
\begin{table}[h]
\caption{
Model simulation parameters with some fixed values.
}
\label{tab.simu-param}
\begin{ruledtabular}
\begin{tabular}{lc}
 Parameters&
\\
\hline
 $T=298K$&
 room temperature\\
 $\sigma_0=90/L^2$ &
 macroion surface charge\\
 $Z=1$&
 microion valence\\
 $a =4.25$ \AA\ &
 microion diameter\\
 $l_{B}=1.68a =7.14$ \AA\ &
 Bjerrum length\\
 $L=22 a $&
 $(x,y)$-box length\\
 $\tau=75 a $&
 $z$-box length\\
 $N_+$&
 number of PCs\\
 $N_-$&
 number of PAs\\
 $N_{PE}=N_+ + N_-$&
 total number of PEs\\
 $N_m$&
 number of monomers per chain\\
 $\chi_{vdw}$&
 strength of the specific VDW attraction
\end{tabular}
\end{ruledtabular}
\end{table}
\begin{table}
\caption{
System parameters. The number of counterions (cations and anions) ensuring
the overall electroneutrality of the system is not indicated.
}
\label{tab.simu-runs}
\begin{ruledtabular}
\begin{tabular}{lcccc}
 System&
 $N_{PE}$&
 $N_+$&
 $N_-$&
 $N_m$
\\
\hline
 $A$&
 $20$&
 $20$&
 $0$&
 $10$\\
 $B$&
 $10$&
 $10$&
 $0$&
 $20$\\
 $C$&
 $40$&
 $20$&
 $20$&
 $10$\\
 $D$&
 $20$&
 $10$&
 $10$&
 $20$\\
 $E$&
 $80$&
 $40$&
 $40$&
 $10$\\
 $F$&
 $40$&
 $20$&
 $20$&
 $20$\\
\end{tabular}
\end{ruledtabular}
\end{table}

\newpage
\begin{center}
{\large FIGURE CAPTIONS}
\end{center}

\begin{enumerate}
\item 
Profiles of PC monomer-density $n_+(z)$ at different $\chi_{vdw}$ couplings
(systems $A$ and $B$). The inset corresponds to $\chi_{vdw}=5$ where the two 
curves ($N_m=10$ and 20) are nearly indistinguishable.

\item 
Fraction $\bar{N}_+(z)/(N_+N_m)$ of adsorbed PC monomers at different $\chi_{vdw}$ couplings
(systems $A$ and $B$).

\item 
Net fluid charge $\sigma(z)$ at different $\chi_{vdw}$
couplings (systems $A$ and $B$).

\item 
Typical equilibrium configurations for PC chains adsorbed onto an oppositely
charged planar macroion (systems $A$ and $B$). 
The little counterions are omitted for clarity.
(a) $\chi_{vdw}=0$, $N_m=10$ 
(b) $\chi_{vdw}=0$, $N_m=20$ 
(c) $\chi_{vdw}=5$, $N_m=10$ 
(d) $\chi_{vdw}=5$, $N_m=20$.

\item 
Profiles of monomer density $n_{\pm}(z)$ for oppositely charged polyelectrolytes (systems $C$ and $D$).
(a) $\chi_{vdw}=0$.
(b) $\chi_{vdw}=5$.

\item 
Typical equilibrium configurations for the adsorption of oppositely
charged PE chains (systems $C$ and $D$) onto a planar macroion. 
The polycations are in white and the polyanions in red.
The little ions are omitted for clarity.
(a) $\chi_{vdw}=0$, $N_m=10$ 
(b) $\chi_{vdw}=0$, $N_m=20$ 
(c) $\chi_{vdw}=5$, $N_m=10$ 
(d) $\chi_{vdw}=5$, $N_m=20$. 

\item 
Fraction of adsorbed monomers $\bar{N}_{\pm}(z)$ for oppositely charged polyelectrolytes (systems $C$ and $D$).
(a) $\chi_{vdw}=0$.
(b) $\chi_{vdw}=5$.

\item 
Net fluid charge $\sigma(z)$ at different $\chi_{vdw}$
couplings (systems $C$ and $D$).

\item 
Profiles of monomer density $n_{\pm}(z)$ for oppositely charged polyelectrolytes 
(systems $E$ and $F$).
(a) $\chi_{vdw}=0$.
(b) $\chi_{vdw}=5$.

\item 
Typical equilibrium configurations for the adsorption of oppositely
charged PE chains (systems $E$ and $F$) onto a planar macroion. 
The polycations are in white and the polyanions in red.
The little ions are omitted for clarity.
(a) $\chi_{vdw}=0$, $N_m=10$ 
(b) $\chi_{vdw}=0$, $N_m=20$ 
(c) $\chi_{vdw}=5$, $N_m=10$ 
(d) $\chi_{vdw}=5$, $N_m=20$. 

\item 
Net fluid charge $\sigma(z)$ at different $\chi_{vdw}$
couplings (systems $E$ and $F$).
\end{enumerate}
%
%
\newpage

\newpage

\newpage

\newpage

\newpage

\newpage

\newpage

\newpage

\newpage

\newpage

\end{document}